\documentclass[11pt]{article}
\usepackage{graphicx}

\begin{document}
%

\begin{center}
{\large \bf Neutrinos as a probe of dark-matter particles}

\vskip.5cm

W-Y. Pauchy Hwang\footnote{Correspondence Author;
 Email: wyhwang@phys.ntu.edu.tw; arXiv:xxxx (hep-ph, to be submitted)} \\
{\em Asia Pacific Organization for Cosmology and Particle Astrophysics, \\
Institute of Astrophysics, Center for Theoretical Sciences,\\
and Department of Physics, National Taiwan University,
     Taipei 106, Taiwan}
\vskip.2cm


{\small(July 27, 2012)}
\end{center}

\begin{abstract}
We try to envision that there might be a dark-matter world and
neutrinos, especially the right-handed ones, might be coupled directly
with dark-matter particles in the dark-matter world. The candidate
model would be the extended Standard Model based on $SU_c(3) \times
SU_L(2) \times U(1) \times SU_f(3) \times SU_R(2)$, with the search of
the detailed version through the aid of the two working rules, "Dirac
similarity principle" and "minimum Higgs hypothesis".

\bigskip

{\parindent=0pt PACS Indices: 12.60.-i (Models beyond the standard
model); 98.80.Bp (Origin and formation of the Universe); 12.10.-g
(Unified field theories and models).}
\end{abstract}

\section{Why are neutrinos so interesting?}

Neutrinos have masses, the tiny masses outside the range of the masses of
the quarks and charged leptons. This is one of the most important experimental
results over the last thirty years. What are the theoretical implications of
this fact?

If we take some minimal extension of the Standard-Model group, that
neutrinos have tiny masses could be taken as a signature that there is a
heavy extra $Z^{\prime 0}$, so that a new Higgs doublet should exist. This
extra $Z^{\prime 0}$ then requires the new "remote" Higgs doublet\cite{Hwang}.
This Higgs doublet also generates the tiny neutrino masses. This is the
$SU_c(3)\times SU_L(2)\times U(1) \times U(1)$ extension. Is this true?

However, there is room left for something very interesting. Remember that
the right-handed neutrinos never enter in the construction of the minimal
Standard Model \cite{Books}. The message that the right-handed neutrinos
seem to be "unwanted" could be telling us something. Now, the fact that
neutrinos have tiny masses suggests that "more naturally" they would be
four-component Dirac particles, and maybe unlikely to be the two-component
Majorana particles.

Now, the right-handed neutrinos are "unwanted" in the minimal Standard
Model, leaving the room so that they could form some multiplet under a
new gauge group (beyond the minimal Standard Model). In fact, we have
some candidate from the symmetries - the family symmetry that there are three
generations in the building blocks of ordinary matter, and so far only three.
We have seen this fact, but we don't know why - let's speculate that it could
be the story associated with the dark-matter world.

Thus, it arises naturally the so-called family gauge theory \cite{Family}. Note
that the right-handed neutrinos do not appear in the minimal Standard Model.
Indeed, we could make a massive $SU_f(3)$ gauge theory completely independent of
the minimal Standard Model, including the particle content. We
could treat $(\nu_{\tau R},\, \nu_{\mu R},\, \nu_{e R})$ as a triplet under
this $SU_f(3)$ - so to give rise to a family gauge theory. Because the
anomaly might not hurt, we could drop the right-handed labels from
the neutrinos. This completes the derivation of the family gauge theory
\cite{Family}. The $SU_f(3)$ is by definition the massive gauge theory
all the involved particles are dark-matter particles in the dark-matter
world.

Basically, we could try to combine the minimal Standard Model
$SU_c(3)\times SU_L(2) \times U(1)$ with $SU_f(3)$, with $(\nu_{\tau R},
\,\nu_{\mu R},\, \nu_{eR})$ the basic $SU_f(3)$ triplet. Here $SU_f(3)$
has an orthogonal neutrino multiplet since the right-handed neutrinos do
not enter at all the minimal Standard Model. In this way, we obtain the
$SU_c(3) \times SU_L(2) \times U(1) \times SU_f(3)$ minimal model. Or,
the right-handed indices could be dropped altogether in the
family group, just like the other $SU_c(3)$ combining with $SU_L(2)
\times U(1)$ as far as anomalies are concerned.

In this case \cite{Family}, the three family calls for $SU_f(3)$ and to make
the gauge bosons all massive the minimum choice would be a pair of complex Higgs
triplets - apparently a kind of broken gauge symmetry. The scenario is such that
the eight gauge bosons and the four left-over Higgs particles all have masses
greater than a few $TeV's$ (i.e. above the LHC energies). The reason why they
all should be massive is that they can enter the loop diagrams in the
ordinary-matter world (i.e., the world initially described by the minimal Standard
Model). At this point, it seems that there is some sort of "minimum Higgs
hypothesis", after forty years of searching for the Higgs particles. Under the
"minimum Higgs hypothesis", the structure of the
underlying Higgs mechanism is pretty much determined. Then, neutrinos
acquire their masses, to the leading order, with the aid of both the Higgs
triplets. In addition, the loop diagrams involving the gauge bosons (familons)
also contribute to neutrino masses. The dark-matter origins of neutrino masses
may explain why they are so tiny.

We know that there are three generations of quarks and leptons
but don't know exactly why. This may be a symmetry that we have already
"seen", but not in details. In view of the existing structure of the
minimal Standard Model, we suspect that most symmetries might be
realized in the form of gauge theories (our own prejudice), in this case
a family gauge theory \cite{Family}. On other hand, the missing
right-handed sector is always a mystery to many of us. So, the
next option is to employ the notion originated by Pati and Salam
\cite{Salam} that the left-right symmetry is restored
at some even higher energy. Again, the Higgs sector is determined via
"minimum Higgs hypothesis"; otherwise, there are too many choices as
originally proposed \cite{Salam}, even if the Higgs sector is concerned.

We could move one step forward. We see that the three generations
are already there (even though we have not seen the feeble interactions so
far). Judging from the energies which we could reach at LHC, we could set
the mass scale at a few $TeV's$ (for familons and family Higgs). On the
other hand, until the LHC energies we haven't seen any signature that the
right-handed sector would be back - so setting the scale to at least
hundreds of $TeV's$. In other words, we are talking
about the extended Standard Model based on the group $SU_c(3) \times SU_L(2)
\times U(1) \times SU_f(3) \times SU_R(2)$.

To summarize, we think that, when we talk about jointly the ordinary-matter
world and the dark-matter world, the family symmetry is there (explaining why
there are three generations of fermions) and the missing right-handed sector,
though still not there, would eventually come back (as energies increase).
Everything is written through the Standard-Model way, or in the gauge-field
fashion. It is naively renormalizable (i.e., by power counting). It is clear
that the minimum extended Standard Model would be the extended Standard Model
to be based on the group $SU_c(3) \times SU_L(2) \times U(1) \times
SU_f(3) \times SU_R(2)$. That is our rationale.

Maybe in the following we would like to append two lengthy remarks - first,
on the possible existence of dark-matter galaxies; and, second, on the
validity of the two working rules, "Dirac similarity principle"
and "minimum Higgs hypothesis".

We would be curious about how the dark-matter world looks like, though
it is difficult to verify experimentally. The first question would be: The
dark-matter world, 25 \% of the current Universe (in comparison, only 5 \%
in the ordinary matter), would clusterize to form the dark-matter
galaxies, maybe even before the ordinary-matter galaxies. The dark-matter
galaxies would then play the hosts of (visible) ordinary-matter galaxies,
like our own galaxy, the Milky Way. Note that a dark-matter galaxy is
by our definition a galaxy that does not possess any ordinary strong and
electromagnetic interactions (with our visible ordinary-matter world),
i.e. that consists of dark-matter particles only. This fundamental
question deserves some thoughts, for the structural formation of
our Universe.

In the ordinary-matter world, strong and electromagnetic forces make the
clustering a very different story - they manufacture atoms, molecules, complex
molecules, and chunks of matter, and then the stars and the galaxies; the
so-called "seeded clusterings". The relatively-rapid story up to chunks of
matter is mainly due to the residual electromagnetic and strong forces, rather
than from gravitational forces. On the other hand, the seeds in the
dark-matter world could come from relatively-stable extra-heavy dark-matter
particles (such as familons and family Higgs, depending on how many decay
channels) - one such particle would be equivalent to thousands of
ordinary-matter molecules. Note that, in the ordinary-matter world, the sequence
of atoms-molecules-complex molecules-etc., up to the mass of the $TeV$ level,
yields the "seeds" of the clusterings - the seeded clustering that is
relevant for the time span (in about $1\, Gyr$) of our young Universe in the
ordinary-matter world. In our extended Standard Model, such seeds for
dark-matter galaxies might come from relatively-stable heavy dark-matter particles
(with the mass greater than, e.g., a few $TeV$). They live relatively long in the
absence of decay channels.

The idea of the "seeded" clustering, and its potential importance in galactic
formation and evolution is rather intuitive - hence intuitively acceptable.
However, this idea is at odd with the standard wisdom that the gravitational
force be solely responsible for clusterings - and for galactic formation
as well. To determine which one is correct for the time span of about $1\,
Gyr$ (i.e. the age of our young Universe), serious numerical simulations
may be needed.

On the other hand, it would be too broad to identify which extended Standard
Model, in detail, could be the final choice. With the "minimum Higgs hypothesis"
\cite{Hwang3}, the Higgs sector is essentially fixed once the group
$SU_c(3) \times SU_L(2) \times U(1) \times G$ with $G$ the extension is
fixed. On the other hand, the "Dirac Similarity Principle" helps to fix
the particle contents. With these working rules \cite{Hwang3} that are
in essence used for several decades, the search for the correct extended
Standard Model could be sharpened and thus much easier.

Or, we could construct so many extended Standard Models, even if the gauge
group is fixed to be, e.g., $SU_c(3) \times SU_L(2) \times U(1) \times
SU_f(3) \times SU_R(2)$ -- due to the unknown Higgs multiplets and
due to whether the fermions would be Dirac particles and, furthermore, the
standard-model multiplets would be the best choice. Using "Dirac
similarity principle" and "minimum Higgs hypothesis" \cite{Hwang3}, we
proceed to make choices only those "naively" renormalizable extended
Standard Model(s) and to test its validity.

To push forward the "final" Standard Model, we should have a comprehensive
successful phenomenology \cite{PDG}. On the $SU_c(3) \times SU_L(2) \times
U(1) \times SU_f(3) \times SU_R(2)$ extended Standard Model, the first
goal is to pin down the manifestations of the $SU_f(3)$ gauge sector.
The physics of the neutrino sector gets modified and the $\tau-\mu-e$
universality is in principle no longer there, depending on the strength
of the $SU_f(3)$ coupling. The reason for the neutrino couplings is that
the neutrino is only species in the ordinary-matter world that acts also
as dark matter. And if the $SU_f(3)$ gauge sector is there, its
communication with us (the ordinary matter) is only through the
neutrinos. So, the breakdown of the $\tau-\mu-e$ universality,
albeit it could be very small, is absolutely crucial.

In the said family gauge theory, we introduce a pair of complex Higgs triplets
in order to guarantee that all gauge bosons ("familons") and the remaining four
Higgs particles are massive, say, greater than a few $TeV$. They have to be massive
mainly because, if massless, the loop diagrams involving these dark-matter
particles could become dominant. We also note that the pair of complex Higgs
triplets and the neutrino triplet can form a singlet, the off-diagonal neutrino
mass interaction, providing a natural mechanism for neutrino oscillations.

\section{Where could we see the detailed tests?}

Neutrino oscillations are firmly established, though much more yet
to come; this is where $\nu_e$ suddenly reappears as $\nu_\mu$ or as
$\nu_\tau$, where $\nu_\mu$ as $\nu_\tau$ or as $\nu_e$, and so on.
This is precisely the lepton-flavor-violating interaction as given by
\begin{equation}
i \eta {\bar \Psi} \cdot \{\Phi_+ + \epsilon \Phi_- \} \times \Psi,
\end{equation}
with $\Phi_\pm$ the pair of complex Higgs triplets and $\Psi$ the neutrino
triplet $(\nu_\tau,\, \nu_\mu,\, \nu_e)$. Here the remaining Higgs
particles are massive and the associated vacuum expectation values
are given by $u_\pm$ \cite{Family}. The role of this off-diagonal
interaction, a lepton-flavor-violating interaction, may be underestimated.

\begin{figure}[h]
\centering
\includegraphics[width=4in]{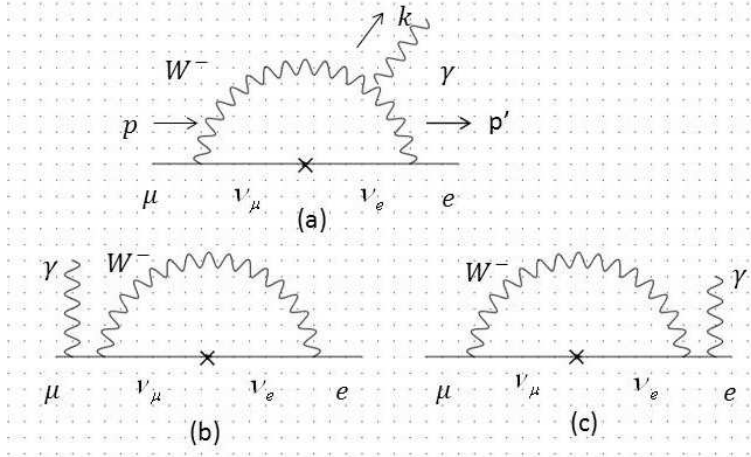}
\caption{The leading diagrams for $\mu \to e + \gamma$.}
\end{figure}

The other important possibilities come from the serious search for
lepton-flavor-violating decays (reactions) of charged leptons, such as
$\mu\to e + \gamma$, $\mu\to 3e$, $\mu + A \to e + A^*$, and so on.
For $\mu \to e + \gamma$ as example, we show in Figs. 1(a), 1(b), and 1(c)
that the off-diagonal interaction is in operation. Using a recent
calculation \cite{Family1}, it is reduced to an effective interaction as
given by

\begin{eqnarray*}
T= 2\cdot {G_F\over \sqrt 2} &\cdot \eta (u_+ + \epsilon u_-)
\cdot (m_1 + m_2)\cdot (+2i){e\over (4\pi)^2}\nonumber\\
&\cdot {\bar u}(p',s') {\gamma\cdot \epsilon\over \sqrt {2k_0}}
(1+\gamma_5) u(p,s).
\end{eqnarray*}
The decay rate is determined by $d\Gamma={d^3p'\over (2\pi)^3}
{d^3k\over (2\pi)^3} (2\pi)^4 \delta^4(p-p'-k) {1\over 2} \sum |T|^2$,
yielding the branching ratio proportional to $O(m^4/m_\mu^4)$.
The tininess of the neutrino masses makes the observation virtually
impossible.

Maybe the other possibility is through the detailed tests of the
$\tau-\mu-e$ universality. Years ago, this universality entered the 
arguments in the $\tau$-mass measurements. We need to re-check the 
arguments in view of the present tiny deviations. In addition, the 
tests of the Standard Model involve the so-called $\rho$ parameter 
- to check the $\mu-e$ universality, we could
compare the reactions induced by $\nu_\mu$ with those induced by
$\nu_e$. It is clear that a new field is opening up.

\section{Episode}

In writing up this note, I have kept in mind that the spiral tail of the Milky
Way is caused by the dark-matter aggregate, or even a dark-matter galaxy, of
four or five times the mass of the Milky Way, and similarly for other spiral
galaxies. In the fronts of astrophysics, further investigations of these
spiral galaxies may eventually offer a lot more in the dark-matter world.

Of course, we should remind ourselves that, in our ordinary-matter world,
those quarks can aggregate in no time, to hadrons, including nuclei, and
the electrons serve to neutralize the charges also
in no time. Then atoms, molecules, complex molecules, and so on. These serve as
the seeds for the clusters, and then stars, and then galaxies, maybe in a time span
of $1\, Gyr$ (i.e., the age of our young Universe). The aggregation caused by
strong and electromagnetic forces is fast enough to help giving rise to galaxies
in a time span of $1\, Gyr$. On the other hand, the seeded clusterings might
proceed with abundance of extra-heavy dark-matter particles such as familons
and family Higgs, all greater than a few $TeV$ and with relatively long
lifetimes (owing to very limited decay channels). So, further simulations on
galactic formation and evolution may yield clues on our problem.

Finally, coming back to the fronts of particle physics, neutrinos,
especially the right-handed neutrinos, might couple to the dark-matter
particles. Any further investigation along this direction would be of
utmost importance. It may shed light on the nature of the dark-matter world.

\section*{Acknowledgments}
This research is supported in part by National Science Council project (NSC
99-2112-M-002-009-MY3). We wish to thank the authors of the following
books \cite{Books} for thorough reviews of the minimal Standard Model.

\end{document}